\documentclass{svmult}

\usepackage{makeidx}         
\usepackage{graphicx}        
\usepackage{multicol}        
\usepackage[bottom]{footmisc}

\makeindex             

\newcommand{\sgn}{\mbox{sgn}}
\newcommand{\eeq}{\end{equation}}

\newcommand{\br}{\mbox{\boldmath $r$}}

\newcommand{\bs}{\mbox{\boldmath $s$}}

\newcommand{\bb}{\mbox{\boldmath $b$}}

\newcommand{\bp}{\mbox{\boldmath $p$}}

\newcommand{\bM}{\mbox{\boldmath $M$}}

\newcommand{\bP}{\mbox{\boldmath $P$}}

\newcommand{\bH}{\mbox{\boldmath $H$}}

\newcommand{\bk}{\mbox{\boldmath $k$}}

\newcommand{\bS}{\mbox{\boldmath $S$}}

\newcommand{\bn}{\mbox{\boldmath $n$}}

\newcommand{\bd}{\mbox{\boldmath $d$}}

\newcommand{\bR}{\mbox{\boldmath $R$}}

\newcommand{\bI}{\mbox{\boldmath $I$}}

\newcommand{\bD}{\mbox{\boldmath $D$}}

\newcommand{\ds}{\displaystyle}
\newcommand{\bw}{\mbox{\boldmath $w$}}

\newcommand{\K}{{\cal K}}
\newcommand{\G}{{\cal G}}
\newcommand{\N}{{\cal N}_0}
\def\R{{\cal R}}
\def\S{{\cal S}}
\def\I{{\cal I}}

\newcommand{\beq}{\begin{equation}}

\begin{document}

\title*{A Stochastic Non-Cooperative Game for Energy Efficiency in Wireless Data Networks}
\titlerunning{A stochastic game in wireless data networks}
\author{Stefano Buzzi\inst{1}\and
H. Vincent Poor\inst{2}\and
Daniela Saturnino\inst{1}}
\institute{Universit\`{a} degli Studi di Cassino \\
03043 Cassino (FR) - ITALY \\
\texttt{\{buzzi, d.saturnino\}@unicas.it}
\and School of Engineering and Applied Science \\
Princeton University \\ Princeton, NJ, 08544, USA \\
\texttt{poor@princeton.edu}\thanks{This paper was supported by the U. S. Air Force Research
Laboratory under Cooperative Agreement No. FA8750-06-1-0252, and by the U. S. Defense Advanced Research Projects
Agency under Grant No. HR0011-06-1-0052.}}
%
%
\maketitle

\begin{abstract}
In this paper the issue of energy efficiency in CDMA wireless data networks is addressed through a game theoretic approach. Building on a recent paper by the first two authors, wherein a non-cooperative game for spreading-code optimization, power control, and receiver design has been proposed to maximize the ratio of data throughput to transmit power for each active user, a stochastic algorithm is here described to perform adaptive implementation of the said non-cooperative game. The proposed solution is based on a combination of RLS-type and LMS-type adaptations, and makes use of readily available measurements. Simulation results show that its performance approaches with satisfactory accuracy that of the non-adaptive game, which requires a much larger amount of prior information.
\end{abstract}

\section{Introduction}

Game theory \cite{gtbook} is a branch of mathematics that has been applied primarily in economics and other social sciences to study the interactions among several autonomous subjects with contrasting interests. More recently, it has been discovered that it can also be used for the design and analysis of communication systems, mostly with application to resource allocation algorithms \cite{gt}, and, in particular, to power control \cite{yates}.
As examples, the reader is referred to \cite{nara1,nara2,SaraydarPhD}. Here, for a multiple access wireless data network,
noncooperative and cooperative games are introduced, wherein each user chooses its  transmit power in order to maximize its own utility, defined as the ratio of the throughput to transmit power.
While the above papers consider the issue of power control assuming that a conventional matched filter is available at the receiver, reference \cite{meshkati} considers the problem of joint linear receiver design and power control so as to maximize the utility of each user. It is shown here that the inclusion of receiver design in the considered game brings remarkable advantages, and, also, results based on the powerful large-system analysis are presented.
More recently, the results of \cite{meshkati} have been extended in \cite{ew2007} to the case in which also each user's spreading code is included in the tunable parameters for utility maximization. The study \cite{ew2007} thus shows that significant performance gains can be obtained through the joint optimization of the spreading code, the transmit power and the receiver filter for each user.

On the other hand, the solutions proposed in \cite{meshkati} and \cite{ew2007}, while providing a general framework for cross-layer resource optimization through a game theoretic approach, describe solutions based on a perfect knowledge of a number of parameters such as the spreading codes, the transmit powers, the propagation channels and the receive filters for all the users. Otherwise stated, the optimization procedure for each user requires a vast amount of prior information not only for the user of interest, but also for all the remaining active users. In this paper, instead, we consider the more practical and challenging situation in which each user performs utility maximization based on the knowledge of its parameters only, i.e. assuming total ignorance of the interference background. This may be the usual scenario in the downlink of a wireless data network, as well as in the uplink of a multicell wireless network, wherein each access point (AP) is disturbed by the interference originating from users served by surrounding AP's.
An adaptively learning algorithm capable of approaching with good accuracy the performance of the non-adaptive game is thus presented, based on a combination of the recursive-least-squares (RLS) and least-mean-squares (LMS) adaptation rules. The proposed algorithm assumes no prior knowledge on the interference background and makes use of readily available measurements.

The rest of this paper is organized as follows. The next section contains some preliminaries and the system model of interest. Section 3 contains a brief review of the non-adaptive game considered in \cite{ew2007}, while the stochastic implementation of the resource allocation algorithm is detailed in Section 4. Section 5 contains extensive simulation results, while, finally concluding remarks are given in Section 6.

\section{Preliminaries and problem statement}
Consider the uplink of a $K$-user synchronous, single-cell, direct-sequence code division multiple access (DS/CDMA) network with processing gain $N$ and subject to flat fading. After chip-matched filtering and sampling at the chip-rate, the $N$-dimensional received data vector, say $\br$, corresponding to one symbol interval, can be written as
\beq
\br=\ds \sum_{k=1}^{K}\sqrt{p_k} h_k b_k \bs_k + \bn \; ,
\label{eq:r}
\eeq
wherein $p_k$ is the transmit power of the $k$-th user\footnote{To simplify subsequent notation, we assume that the transmitted power $p_k$ subsumes also the gain of the transmit and receive antennas.}, $b_k\in \{-1,1\}$ is the information symbol of the $k$-th user, and $h_k$ is the real\footnote{We assume here, for simplicity, a real channel model; generalization to practical channels, with I and Q components, is straightforward.} channel gain between the $k$-th user's transmitter and the access point (AP); the actual value of $h_k$ depends on both the distance of the $k$-th user's terminal from the AP and the channel fading fluctuations. The $N$-dimensional vector $\bs_k$ is the spreading code of the $k$-th user; we assume that the entries of $\bs_k$ are real and that $\bs_k^T \bs_k=\|\bs_k\|^2=1$, with $(\cdot)^T$ denoting transpose. Finally, $\bn$ is the ambient noise vector, which we assume to be a zero-mean white Gaussian random process with covariance matrix $(\N/2) \bI_N$, with $\bI_N$ the identity matrix of order $N$. An alternative and compact representation of (\ref{eq:r}) is given by
\beq
\br=\bS \bP^{1/2}\bH \bb + \bn \; ,
\label{eq:r2}
\eeq
wherein $\bS=[ \bs_1, \ldots, \bs_K]$ is the $N\times K$-dimensional spreading code matrix, $\bP$ and $\bH$ are $K \times K$-dimensional diagonal matrices, whose diagonals are $[p_1, \ldots, p_K]$ and $[h_1, \ldots, h_K]$, respectively, and, finally, $\bb=[b_1, \ldots, b_K]^T$ is the $K$-dimensional vector of the data symbols.

Assume now that each mobile terminal sends its data in packets of $M$ bits, and that it is interested both in having its data received with as small as possible error probability at the AP, and in making  careful use of the energy stored in its battery. Obviously, these are conflicting goals, since error-free reception may be achieved by increasing the received SNR, i.e. by increasing the transmit power, which of course comes at the expense of battery life\footnote{Of course there are many other strategies to lower the data error probability, such as for example the use of error correcting codes, diversity exploitation, and implementation of optimal reception techniques at the receiver. Here, however, we are mainly interested to energy efficient data transmission and power usage, so we consider only the effects of varying the transmit power, the receiver and the spreading code on energy efficiency.}. A useful approach to quantify these conflicting goals is to define the utility of the $k$-th user as the ratio of its throughput, defined as the number of information bits that are received with no error in unit time, to its transmit power \cite{nara1,nara2}, i.e.
\beq
u_k=\ds \frac{T_k}{p_k}\; .
\label{eq:utility}
\eeq
Note that $u_k$ is measured in bit/Joule, i.e. it represents the number of successful bit transmissions that can be made for each Joule of energy drained from the battery.
Denoting by $R$ the common rate of the network (extension to the case in which each user transmits with its own rate $R_k$ is quite simple) and assuming that each packet of $M$ symbols contains $L$ information symbols and $M-L$ overhead symbols, reserved, e.g., for channel estimation and/or parity checks, the throughput $T_k$ can be expressed as
\beq
T_k=\ds R \frac{L}{M} P_k
\label{eq:Tk}
\eeq
wherein $P_k$ denotes the probability that a packet from the $k$-th user is received error-free. In the considered DS/CDMA setting, the term $P_k$ depends formally on a number of parameters such as the spreading codes of all the users and the diagonal entries of the matrices $\bP$ and $\bH$, as well as on the strength of the used error correcting codes. However, a customary approach is to model the multiple access interference as a Gaussian random process, and assume that $P_k$ is an increasing function of the $k$-th user's Signal-to-Interference plus Noise-Ratio (SINR) $\gamma_k$, which is naturally the case in many practical situations.

Recall that, for the case in which a linear receiver is used to detect the data symbol $b_k$, according, i.e., to the decision rule
\beq
\widehat{b}_k=\mbox{sign}\left[\bd_k^T \br\right] \; ,
\label{eq:decrule}
\eeq
with $\widehat{b}_k$ the estimate of $b_k$ and $\bd_k$ the $N$-dimensional vector representing the receive filter for the user $k$, it is easily seen that the SINR $\gamma_k$ can be written as
\beq
\gamma_k=\ds \frac{p_k h_k^2 (\bd_k^T \bs_k)^2}{\frac{\N}{2}\|\bd_k\|^2 + \ds \sum_{i \neq k} p_i h_i^2
(\bd_k^T \bs_i)^2} \; .
\label{eq:gamma}
\eeq
Of related interest is also the mean square error (MSE) for the user $k$, which, for a linear receiver, is defined as
\beq
{\rm MSE}_k= E \left\{ \left(b_k - \bd_k^T \br \right)2 \right\}=1 + \bd_k^T \bM \bd_k - 2\sqrt{p_k} h_k \bd_k^T
\bs_k \; ,
\label{eq:msek}
\eeq
wherein $E\left\{ \cdot \right\}$ denotes statistical expectation and $\bM=\left(\bS \bH \bP \bH^T \bS^T + \frac{\N}{2} \bI_N\right)$ is the covariance matrix of the data.

\medskip

The exact shape of $P_k(\gamma_k)$ depends  on factors such as the modulation and coding type.
However, in all cases of relevant interest, it is an increasing function of $\gamma_k$ with a sigmoidal shape, and converges to unity as $\gamma_k \rightarrow + \infty$; as an example, for  binary phase-shift-keying (BPSK) modulation coupled with no channel coding, it is easily shown that
\beq
P_k(\gamma_k)=\left[1-Q(\sqrt{2\gamma_k})\right]^M \; ,
\label{eq:psr}
\eeq
with $Q(\cdot)$ the complementary cumulative distribution function of a zero-mean random Gaussian variate with unit variance. A plot of  (\ref{eq:psr}) is shown in Fig. 1 for the case $M=100$.

It should be also noted that substituting (\ref{eq:psr}) into (\ref{eq:Tk}), and, in turn, into (\ref{eq:utility}), leads to a strong incongruence. Indeed, for $p_k \rightarrow 0$, we have $\gamma_k \rightarrow 0$, {\em but} $P_k$ converges to a small but non-zero value (i.e. $2^{-M}$), thus implying that an unboundedly  large utility can be achieved by transmitting with zero power, i.e. not transmitting at all and making blind guesses at the receiver on what data were transmitted. To circumvent this problem, a customary approach \cite{nara2,meshkati} is to replace $P_k$ with an {\em efficiency function}, say $f_k(\gamma_k)$, whose behavior should approximate as close as possible that of $P_k$, except that for $\gamma_k \rightarrow 0$ it is required that $f_k(\gamma_k)= o(\gamma_k)$. The function
$f(\gamma_k)=(1-e^{-\gamma_k})^M$ is a widely accepted substitute for the true probability of correct packet reception, and in the following we will adopt this model\footnote{See Fig. 1 for a comparison between the
Probability $P_k$ and the efficiency function.}.
This efficiency function is increasing and S-shaped, converges to unity as $\gamma_k$ approaches infinity, and has a continuous first order derivative. Note that we have omitted the subscript $``k''$, i.e. we have used the notation $f(\gamma_k)$ in place of $f_k(\gamma_k)$ since we assume that the efficiency function is the same for all the users.

Summing up, substituting (\ref{eq:Tk}) into (\ref{eq:utility}) and replacing the probability $P_k$ with the above defined efficiency function, we obtain the following expression for the $k$-th user's utility:
\beq
u_k=R \ds \frac{L}{M} \frac{f(\gamma_k)}{p_k} \; , \quad \forall k=1, \ldots, K \; .
\label{eq:utility2}
\eeq

\begin{figure}
\centering
\includegraphics[height=7cm,width=9.5cm]{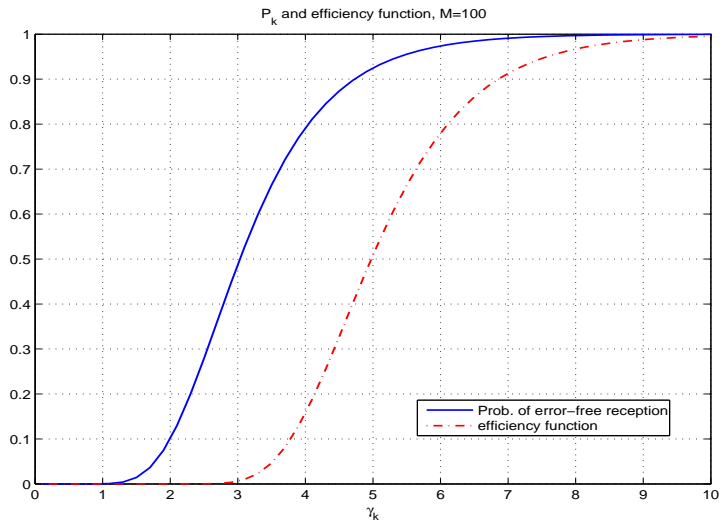}
\caption{Comparison of probability of error-free packet reception and efficiency function versus receive SINR and for packet size $M=100$. Note the S-shape of both functions.}
\label{figampiezza}
\end{figure}

\section{The non-cooperative game with linear receivers}

Based on the utility definition (\ref{eq:utility2}), it is natural to wonder how each user may maximize its utility, how this maximization affects utilities achieved by other users, and, also, if a stable equilibrium point does exist.
These questions have been answered in recent papers by resorting to the tools of game theory. As an example, non-cooperative scenarios wherein mobile users are allowed to vary their transmit power only have been considered in \cite{nara1,nara2,SaraydarPhD};  in  \cite{meshkati}, instead, such an approach has been extended to the cross layer scenario in which each user may vary its power and its uplink  linear receiver, while, more recently, reference \cite{ew2007} has considered the case in which each user is able to tune the transmit power, the uplink linear receiver and the adopted spreading code.

Formally, the game $\G$ considered in \cite{ew2007} can be described as the triplet $\G=\left[\K, \left\{\S_k\right\}, \left\{u_k\right\} \right]$, wherein $\K= \left\{1, 2, \ldots, K\right\}$ is the set of active users participating in the game,
$u_k$ is the $k$-th user's utility defined in (\ref{eq:utility2}), and
\beq
\S_k=[0, P_{k,\max}] \times \R^N \times \R_1^N \; ,
\label{eq:strategy}
\eeq
is the set of possible actions (strategies) that user $k$ can take. It is seen that $\S_k$ is written as the Cartesian product of three different sets, and indeed $[0, P_{k, \max}]$ is the range of available transmit powers for the $k$-th user (note that $P_{k, \max}$ is the maximum allowed transmit power for user $k$), $\R^N$, with $\R$ the real line, defines the set of all possible linear receive filters, and, finally,
$$
\R_1^N= \left\{\bd \in \R^N \; : \; \bd^T \bd=1 \right\} \; ,
$$
defines the set of the allowed spreading codes\footnote{Here we assume that the spreading codes have real entries; the problem of utility maximization with reasonable complexity for the case of discrete-valued entries is a challenging issue that will be considered in the future.} for user $k$.

Summing up, the proposed noncooperative game can be cast as the following maximization problem
\beq
\ds \max_{\S_k} u_k = \max_{p_k, \bd_k, \bs_k} u_k(p_k, \bd_k, \bs_k ) \; , \quad
\forall k=1, \ldots, K \; .
\label{eq:game}
\eeq
Given  (\ref{eq:utility2}), the above maximization can be also written as
\beq
\ds \max_{p_k, \bd_k, \bs_k} \frac{f(\gamma_k(p_k, \bd_k, \bs_k))}{p_k} \; , \quad
\forall k=1, \ldots, K \; .
\eeq
Moreover, since the efficiency function is monotone and non-decreasing, we also have
\beq
\ds \max_{p_k, \bd_k, \bs_k} \frac{f(\gamma_k(p_k, \bd_k, \bs_k))}{p_k}= \max_{p_k}  \frac{f\left(\ds \max_{\bd_k, \bs_k}\gamma_k(p_k, \bd_k, \bs_k)\right)}{p_k} \; ,
\label{eq:deriv}
\eeq
i.e. we can first take care of SINR maximization with respect to spreading codes and linear receivers, and then focus on maximization of the resulting utility with respect to transmit power.

Letting $(\cdot)^+$ denoting Moore-Penrose pseudoinverse, the following result is reported in \cite{ew2007}.

\noindent
 {\bf Proposition:}
{\em The non-cooperative game defined in (\ref{eq:game}) admits a unique Nash equilibrium point $(p_k^*, \bd_k^*, \bs_k^*)$,
for $k=1, \ldots, K$, wherein
\begin{itemize}
\item[-]
$\bs^*_k$ and $\bd^*_k$ are the unique fixed stable $k$-th user
spreading code and receive filter\footnote{Actually the linear receive filter is unique up to a positive scaling factor.} resulting from iterations
\beq
\begin{array}{lll}
\bd_i=\sqrt{p_i} h_i \left(\bS \bH \bP \bH^T \bS^T + \frac{\N}{2} \bI_N\right)^{-1} \bs_i   & \; \forall i=1, \ldots, K \\
\bs_i=\sqrt{p_i} h_i \left(p_i h_i^2 \bD \bD^T + \mu_i \bI_N \right)^{+} \bd_i  & \; \forall i=1, \ldots, K
\end{array}
\label{eq:iterazioni}
\eeq
with $\mu_i$ such that $\|\bs_i\|^2=1$, and $\bD=[\bs_1, \ldots, \bs_K]$. Denote by $\gamma_k^*$ the corresponding SINR.
\item[-]
$p_k^*=\min \{\bar{p}_k, P_{k, \max} \}$, with $\bar{p}_k$ the $k$-th user transmit power such that the $k$-th user maximum SINR $\gamma_k^*$ equals $\bar{\gamma}$, i.e. the unique solution of the equation $f(\gamma)=\gamma f'(\gamma)$, with $f'(\gamma)$ the derivative of $f(\gamma)$.
\end{itemize}}
\noindent

In practice, this result states that the non-cooperative game (\ref{eq:game}) admits a unique Nash equilibrium, that can be reached as follows. First of all, the equation $f(\gamma)=\gamma f'(\gamma)$ is to be solved in order to determine its unique solution $\bar{\gamma}$; then, an iterative procedure starts wherein the system alternates between these two phases:
\begin{itemize}
\item[a.] Given the transmit powers, each user adjusts its spreading code and receive filter through iterations (\ref{eq:iterazioni}) until an equilibrium is reached;
\item[b.] Given the spreading codes and uplink receivers, each user tunes its transmit power so that its own SIR equals $\bar{\gamma}$. Denoting by $\bp=[p_1, \ldots, p_K]$ the users' power vector, and by $\bI(\bp)$ the $K$-dimensional vector whose $k$-th entry $\bI_k(\bp)$ is written as
\beq
\bI_k(\bp)= \ds \frac{\bar{\gamma}}{h_k^2(\bd_k\bs_k)^2} \left(\frac{\N}{2} \|\bd_k\|^2 + \ds \sum_{i \neq k}
p_i h_i^2 (\bd_k \bs_i)^2\right)\; ,
\label{eq:Ikp}
\eeq
the transmit power vector $\bp$ is the unique fixed stable point of the iteration \cite{yates}
\beq
p_k= \left\{ \begin{array}{lll} \bI_k(\bp)\; , & \qquad & {\rm for} \; \bI_k(\bp)\leq P_{k,\max}\; , \\
P_{k,\max}\; , & \qquad & {\rm for} \; \bI_k(\bp)> P_{k,\max}\; ,\end{array} \right. ,
\label{eq:yates}
\eeq
for all $k=1, \ldots, K$.
\end{itemize}
Steps a. and b. are to be repeated until convergence is reached. It is crucial to note that computation of the equilibrium transmit power, spreading code and linear receiver for each user needs a lot of prior information. In particular, it is seen from eq. (\ref{eq:iterazioni}) that computation of the $k$-th user receiver requires knowledge of the spreading codes, transmit powers and channel gains for all the active users, while computation of the $k$-th user spreading code requires knowledge of $\bD$, i.e. the matrix of the uplink receivers for all the users. Likewise, implementation of iterations (\ref{eq:yates}) also requires the same vast amount of prior information. Our next goal is thus to propose a stochastic resource allocation algorithm that alleviates the need for such prior knowledge, and that is amenable to a decentralized implementation, wherein each user may allocate its own resources based only on knowledge that is readily available, and with total ignorance on the interference background.

\section{Adaptive energy efficient resource allocation}
In order to illustrate the adaptive implementation of the non-cooperative game, a slight change of notation is needed. Indeed, since any adaptive algorithm relies on several data observations in consecutive symbol intervals, we cannot restrict any longer our attention to one symbol interval only, and
we thus denote by $\br(n)$
the $N$-dimensional received data vector in the $n$-th bit interval, i.e.
\beq
\br(n)=\ds \sum_{k=1}^{K}\sqrt{p_k(n)} h_k b_k(n) \bs_k(n) + \bw(n) \; .
\label{eq:rn}
\eeq
Eq. (\ref{eq:rn}) differs from Eq. (\ref{eq:r}) in that a temporal index has been added to some parameters, to underline their time-varying nature: as an example,  $p_k(n)$ and $\bs_k(n)$ are the transmit power and the spreading code of the $k$-th user in the $n$-th symbol interval. Note also that the channel gain does not depend on $n$, i.e. we are implicitly assuming a slow fading channel, even though generalization to the case of slowly time-varying channels is quite straightforward.
In order to obtain an adaptive implementation of the utility maximizing algorithm for the generic $k$-th user, first  we focus on iterations (\ref{eq:iterazioni}); as specified in \cite{ew2007,ulukusyener}, the unique fixed stable point of these iterations achieves the global minimum for the total mean square error (TMSE), which is given by
\beq
{\rm TMSE}= \ds \sum_{k=1}^K {\rm MSE}_k\; ,
\label{eq:tmse}
\eeq
with ${\rm MSE}_k$ defined as in (\ref{eq:msek}). It can be shown, indeed, that minimization of the TMSE leads to a Pareto-optimal solution to the problem of maximizing the SINR for each user. On the other hand, an alternative approach is to consider the case in which each user tries to minimize its own MSE with respect to its spreading code and linear receiver. The $k$-th user MSE can be shown to be written as
\beq
{\rm MSE}_k= 1+ \bd_k^T \left(p_k h_k^2 \bs_k \bs_k^T + \bM_k \right)\bd_k - 2 \sqrt{p_k} h_k \bd_k^t \bs_k \; ,
\label{eq:msek2}
\eeq
with $\bM_k= \bM - p_k h_k^2 \bs_k \bs_k^T$. Minimization of Eq. (\ref{eq:msek2}) with respect to $\bd_k$ and $\bs_k$, under the constraint $\|\bs_k\|^2=1$, yields the iterations
\beq
\begin{array}{lll}
\bd_i=\sqrt{p_i} h_i \left(\bS \bH \bP \bH^T \bS^T + \frac{\N}{2} \bI_N\right)^{-1}\!\! \bs_i   & \; \forall i=1, \ldots, K \\
\bs_i=\sqrt{p_i} h_i \left(p_i h_i^2 \bd_i \bd_i^T + \mu_i \bI_N \right)^{-1} \bd_i  & \; \forall i=1, \ldots, K
\end{array}
\label{eq:iterazioni2}
\eeq
In general, minimization of the TMSE is not equivalent to individual minimization of the MSE of each user; however, in our scenario, i.e. in the case of a single-path fading channel, the two approaches can be shown to be equivalent \cite{honig}. As a consequence, we can state that the fixed point of iterations (\ref{eq:iterazioni}) coincides with that of iterations (\ref{eq:iterazioni2}). Note however that, despite such equivalence, the spreading code update for each user in (\ref{eq:iterazioni2}) depends only on parameters of the user itself, and does not require any knowledge on the interference background. The receiver updates in (\ref{eq:iterazioni}) and (\ref{eq:iterazioni2}) are the same, and indeed they coincide with the MMSE receiver.
Accordingly, since the utility maximizing linear receiver is the MMSE filter, we start resorting to the well-known recursive-least-squares (RLS) implementation of this receiver. Letting $\bR(0)= \epsilon \bI_N$, with $\epsilon$ a small positive constant, letting $\lambda$ be a close-to-unity scalar constant, assuming that the receiver has knowledge of the information symbols $b_k(1), \ldots, b_k(T)$, and denoting by $\bd_k(n)$ the estimate of the linear receiver filter for the $k$-th user in the $n$-th symbol interval, the following iterations can be considered
\beq
\begin{array}{ll}
\bk(n)=\ds \frac{\bR^{-1}(n-1) \br(n)}
{\lambda + \br^T(n) \bR^{-1}(n-1) \br(n)}  \; ,\\
\bR^{-1}(n)= \ds \frac{1}{\lambda}
\left[ \bR^{-1}(n-1) - \bk(n) \br^T(n) \bR^{-1}(n-1)
\right] \; , \\
e_k(n)=\bd_k^T(n-1) \br(n) -b_k(n)\; , \\
\bd_k(n)=\bd_k(n-1) - e_k(n) \bk(n) \; .
\end{array}
\label{eq:rls}
\eeq
The last line in (\ref{eq:rls}) represents the update equation
for the detection vector $\bd_k(\cdot)$. Note that this equation,
in turn,
depends on the error vector $e_k(n)$, which, for $n\leq T$, can be
built based on the knowledge of the training symbol $b_k(n)$.
Once the training phase is over, real data detection takes place
and the error  in the third line
of (\ref{eq:rls}) is computed according to the equation
\beq
e_k(n)=\ds   \bd_k^T(n-1)
\br(n) - \sgn \left[\bd_k^T(n-1) \br(n) \right]\; .
\label{eq:ddmode}
\eeq
Given its own receive filter $\bd_k(n)$, the $k$-th user can then modify its spreading code according to the second line of eq. (\ref{eq:iterazioni2}), i.e.:
\beq
\bs_k(n+1)=\sqrt{p_k(n)} h_k \left( p_k(n) h_k^2 \bd_k(n) \bd_k^T(n) + \mu_k(n) \bI_N\right)^{-1} \bd_k(n) \; ,
\label{eq:spreadingcode}
\eeq
with $\mu_k(n)$ a constant such that $\bs_k^T(n) \bs_k(n)=1$. Note that the update in (\ref{eq:spreadingcode}) only requires parameters of the $k$-th user, thus implying that no knowledge on the interference background is needed.
Finally, we have to consider the tuning of the transmit power so that each user may achieve its target SINR $\bar{\gamma}$. This is a classical stochastic power control problem that has been treated, for instance, in \cite{standard}. A possible solution is to consider a least-mean-squares (LMS) update of the trasmit power according to the rule
\beq
\begin{array}{lll}
p_k(n+1)&=&(1-\rho) p_k(n-1) + \rho I_k(n) \; ,\\
p_k(n+1)&=& \min(p_k(n+1), P_{k,{\rm max}}) \; .
\label{eq:powerupdate}
\end{array}
\eeq
In the above equations, the step-size $\rho$ is a close-to-zero positive constant and $\I_k(n)$ is a stochastic approximation of the $k$-th entry of the vector $\bI(\bp)$, and is expressed as
\beq
I_k(n)=\ds \frac{\bar{\gamma}}{h_k (\bd_k^T(n) \bs_k(n))^2}
\left[(\bd_k^T(n) \br(n) )^2 - p_k(n) h_k^2 (\bd_k^T(n) \bs_k(n))^2 \right] \;
\label{eq:Ik}
\eeq
Note that also the update (\ref{eq:powerupdate}) does not require any knowledge on the interference.
To summarize, the algorithm proceeds as follows: for $n\leq T$, only the RLS update (\ref{eq:rls}) is performed; then, for each $n>T$, the algorithm performs the  updates in (\ref{eq:rls}),  (\ref{eq:spreadingcode}) and, finally,  (\ref{eq:powerupdate}). In particular, note that the power update is made in parallel to the spreading code update and receiver update, i.e. without waiting for convergence of the RLS-based adaptive implementation of the MMSE receiver.

Although a theoretical convergence study of this algorithm is definitely worth being undertaken, it is out of the scope of this paper; in the next section, we will discuss the results of extensive computer simulations that will show the excellent behavior of the outlined procedure.

\begin{figure}[t]
\centerline{\hbox{\includegraphics[height=7cm,width=9.5cm]{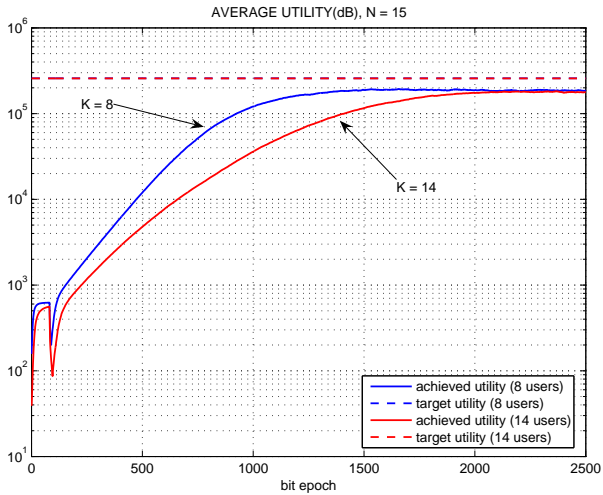}}}
\caption{Achieved average utility versus time. }
\end{figure}

\begin{figure}[!tb]
\centerline{\hbox{\includegraphics[height=7cm,width=9.5cm]{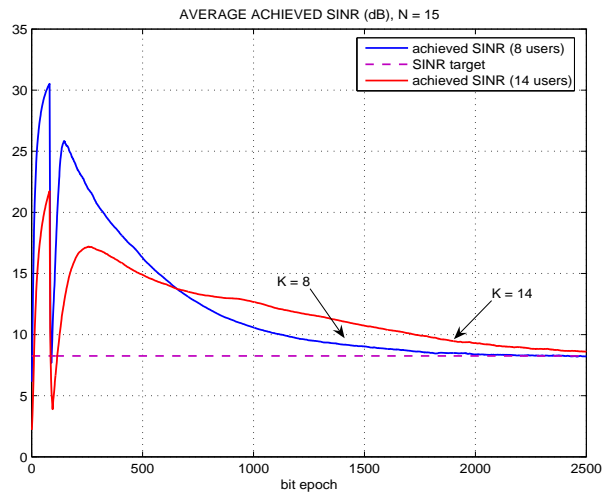}}}
\caption{Achieved average SINR versus time.}
\end{figure}

\begin{figure}[!tb]
\centerline{\hbox{\includegraphics[height=7cm,width=9.5cm]{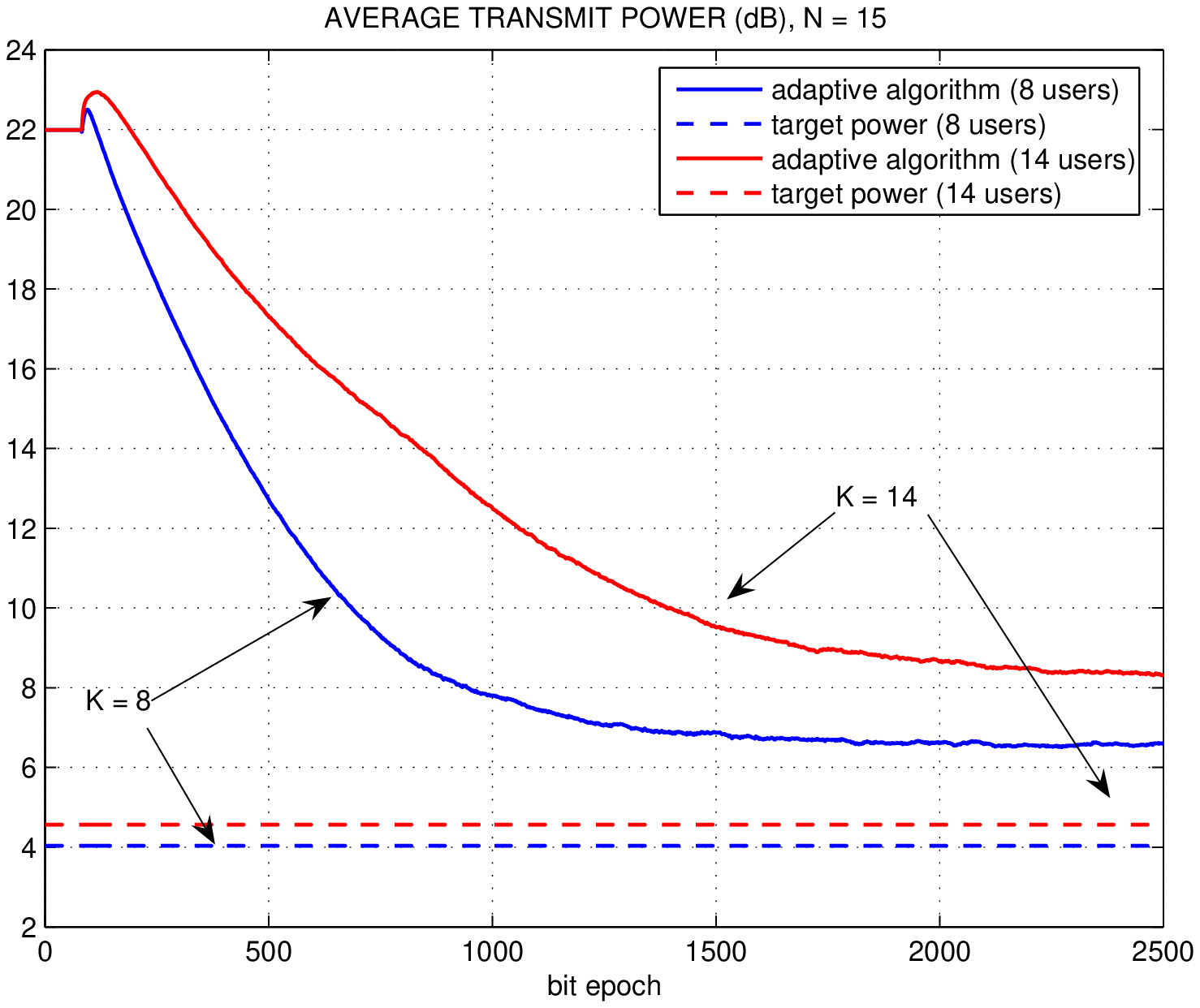}}}
\caption{Average transmit power versus time.}
\end{figure}

\section{Numerical Results}
We contrast here the performance of the non-adaptive game discussed in \cite{ew2007} with its adaptive implementation  proposed here.
We consider an uplink DS/CDMA system with processing gain $N=15$, and assume that the packet length is $M=120$.
for this value of $M$ the equation $f(\gamma)=\gamma f'(\gamma)$ can be shown to admit the solution $\bar{\gamma}=6.689 = 8.25$dB.
A single-cell system is considered, wherein users may have random positions with a distance from the AP ranging from 10m to 500m. The channel coefficient $h_k$ for the generic $k$-th user is assumed to be Rayleigh distributed with mean equal to $d_k^{-2}$, with $d_k$ being the distance of user $k$ from the AP\footnote{Note that we are here assuming that the power path losses are proportional to the fourth power of the path length, which is reasonable in urban cellular environments.}. We take the ambient noise level to be $\N=10^{-5}$W/Hz, while the maximum allowed power $P_{k,\max}$ is $25$dB. We present the results of averaging over $1000$ independent realizations for the users locations, fading channel coefficients and starting set of spreading codes. More precisely, for each iteration we randomly generate an $N \times K$-dimensional spreading code matrix with entries in the set $\left\{-1/\sqrt{N}, 1/\sqrt{N}\right\}$; this matrix is then used as the starting point for the considered games. We consider the case in which $T=80$ training symbols are used, while in eq. (\ref{eq:powerupdate}) the step size $\rho=.01$ has been taken.
Figs. 2 - 4 report the time-evolution of the achieved average utility (measured in bit/Joule), the average achieved SINR and the average transmit power, respectively, for both the cases in which $K=8$ and $K=14$. It is seen that after about one thousand iterations the adaptive algorithm approximate with satisfactory accuracy the benchmark scenario that a non-adaptive game is played as in \cite{ew2007}. In particular, while the target SINR and the achieved utility are quite close to their target values, it is seen from Fig. 4 that the average transmit power is about 3dB larger than in the non-adaptive case; such a loss is not at all surprising, since it is well-known that adaptive algorithms have a steady-state error, and that their performance may only approach that of their non-adaptive counterparts.

In order to test the tracking properties of the proposed algorithm, we also consider a dynamic scenario with an initial number of users $K=8$, and with two additional users entering the channel at time epochs $n=1000$ and $n=1700$. The results are reported in Figs 5 - 7. Results clearly show that the algorithm is capable of coping with changes in the interference background.

\begin{figure}[!tb]
\centerline{\hbox{\includegraphics[height=7cm,width=9.5cm]{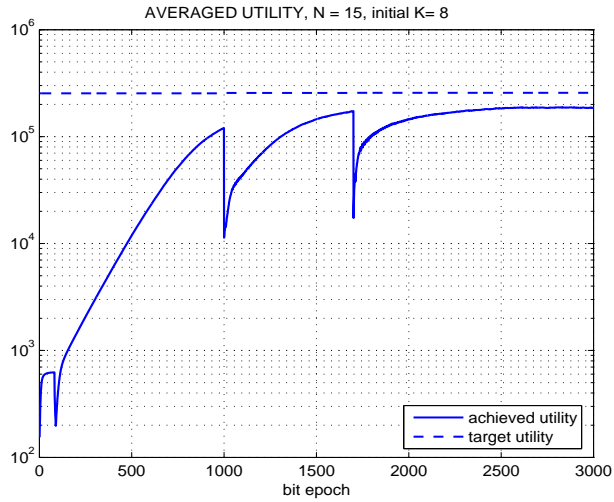}}}
\caption{Dynamic environment: Average achieved utility versus time.}
\end{figure}

\begin{figure}[!tb]
\centerline{\hbox{\includegraphics[height=7cm,width=9.5cm]{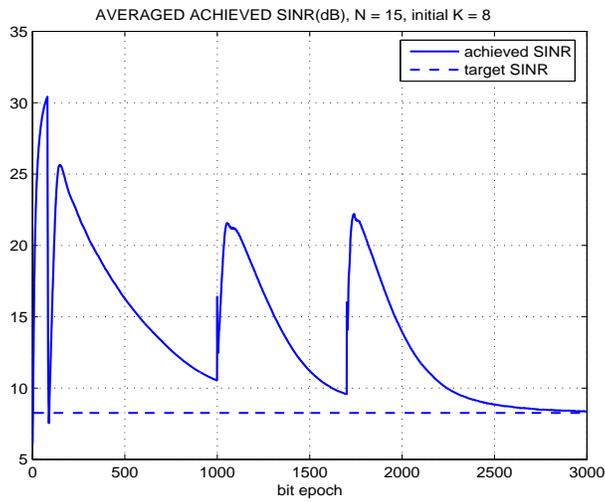}}}
\caption{Dynamic environment: Average achieved SINR versus time.}
\end{figure}

\begin{figure}[!tb]
\centerline{\hbox{\includegraphics[height=7cm,width=9.5cm]{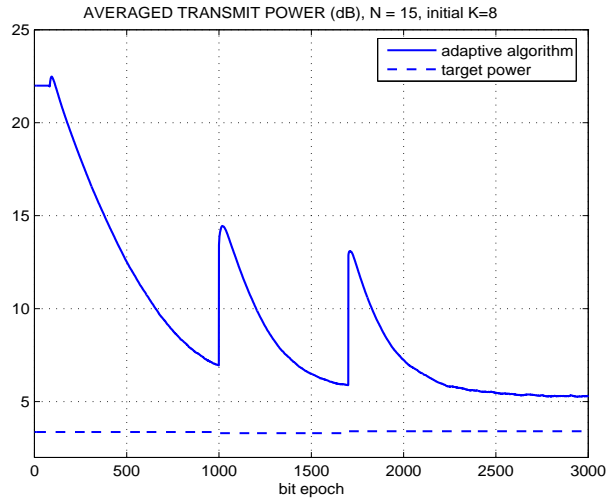}}}
\caption{Dynamic environment: Average transmit power versus time.}
\end{figure}

\section{Conclusion}
In this paper the cross-layer issue of joint stochastic power control, spreading code optimization and receiver design for wireless data networks has been addressed using a game-theoretic framework. Building on \cite{ew2007}, wherein a non-cooperative game for resource allocation has been proposed, we have here considered the issue of adaptive implementation of the resource allocation algorithm, based on readily available measurements and assuming no prior knowledge on the interference background. The result is thus a stochastic algorithm that can be realized in a decentralized fashion, wherein each user just needs knowledge of its own parameters. The performance of the proposed scheme has been validated through computer simulations, which showed that the adaptive implementation achieved a performance quite close to that of the non-adaptive benchmark.
As interesting research topics worth being investigated we mention the theoretical analysis of the convergence properties of the algorithm, and the development and the analysis of adaptive algorithms able to implement the said game without prior knowledge of the fading channel coefficient of the user of interest.

%


\end{document}